\documentclass[runningheads]{llncs}
\usepackage[T1]{fontenc}
\usepackage{graphicx}
\usepackage{url}
\usepackage{multirow}
\usepackage{array}
\begin{document}
\title{Revisiting Network Traffic Analysis: Compatible network flows for ML models}
\titlerunning{Revisiting Network Traffic Analysis}
\author{Jo{\~{a}}o Vitorino\orcidID{0000-0002-4968-3653} \and
Daniela Pinto\orcidID{0009-0000-3003-6694} \and
Eva Maia\orcidID{0000-0002-8075-531X} \and
Ivone Amorim\orcidID{0000-0001-6102-6165} \and
Isabel Pra{\c{c}}a\orcidID{0000-0002-2519-9859}}
\authorrunning{J. Vitorino et al.}
\institute{GECAD, ISEP, Polytechnic of Porto, Rua Dr. António Bernardino de Almeida, 4249-015 Porto, Portugal}
\maketitle
\begin{abstract}
To ensure that Machine Learning (ML) models can perform a robust detection and classification of cyberattacks, it is essential to train them with high-quality datasets with relevant features. However, it can be difficult to accurately represent the complex traffic patterns of an attack, especially in Internet-of-Things (IoT) networks. This paper studies the impact that seemingly similar features created by different network traffic flow exporters can have on the generalization and robustness of ML models. In addition to the original CSV files of the Bot-IoT, IoT-23, and CICIoT23 datasets, the raw network packets of their PCAP files were analysed with the HERA tool, generating new labelled flows and extracting consistent features for new CSV versions. To assess the usefulness of these new flows for intrusion detection, they were compared with the original versions and were used to fine-tune multiple models. Overall, the results indicate that directly analysing and preprocessing PCAP files, instead of just using the commonly available CSV files, enables the computation of more relevant features to train bagging and gradient boosting decision tree ensembles. It is important to continue improving feature extraction and feature selection processes to make different datasets more compatible and enable a trustworthy evaluation and comparison of the ML models used in cybersecurity solutions.
\keywords{Feature extraction \and Feature selection \and Flow exporter \and Packet capture \and Intrusion detection \and Machine learning}
\end{abstract}
\section{Introduction}
\label{sec:introduction}

Securing the Internet-of-Things (IoT) is a never-ending challenge. The complexity of IoT networks, including distributed network topologies, vulnerable devices, and diverse network traffic patterns, makes it difficult to deploy traditional intrusion detection systems~\cite{ENISA2020}. Instead, modern cybersecurity solutions are trying to adapt to evolving cyber threats by integrating artificial intelligence, using Machine Learning (ML) models that are trained to identify suspicious network activity and detect possible zero-day attacks~\cite{Benkhelifa2018}.

To ensure that an ML model is robust against abnormal data, it is essential to train it with high-quality and representative data of an IoT network. Since ML models commonly require the datasets to be in a simple tabular format, it is necessary to prepare Network Traffic Analysis (NTA) datasets by converting network traffic into features that ML models can process~\cite{DIMAURO2021}. Flow exporter tools can preprocess the raw network packets of an IoT network, aggregating them to generate network traffic flows with statistical features that summarise the main characteristics of recent network activity~\cite{Adeleke2022}.

However, it can be difficult to accurately convert and represent the complex traffic patterns of the evolving cyberattacks targeting IoT networks~\cite{VitorinoSoK}. Different flow exporters may implement varying approaches for traffic aggregation, flow filtering, and feature computation, leading to discrepancies in the generated network flows~\cite{Pinto2025}. These discrepancies can cause some portions of a dataset to contain incorrect or biased data, which affects both the generalization and the robustness of ML models, especially in an imbalanced multiclass classification task, which is the case of IoT intrusion detection~\cite{Thakkar2022}.

To address this concern, this paper studies the impact that the features created by different flow exporters can have on the robustness of IoT cyberattack classification models. In addition to the original Comma-Separated Values (CSV) files of the Bot-IoT~\cite{koroniotis_towards_2019}, IoT-23~\cite{sebastian_garcia_2021_4743746}, and CICIoT23~\cite{neto_euclides_2023} datasets, their raw packet capture (PCAP) files were analyzed with a different flow exporter. The developed Holistic nEtwork featuRes Aggregator (HERA) tool~\cite{Pinto2025} was used to generate new labelled flows and extract consistent features, resulting in new CSV versions that were compared with the original versions. To assess the usefulness of the new flows for the training of ML models for NTA and intrusion detection, multiple Random Forest (RF), Extreme Gradient Boosting (XGB), and Light Gradient Boosting Machine (LGBM) models were fine-tuned and evaluated.

This paper is organized into multiple sections. Section 2 provides an overview of the recent scientific developments. Section 3 details the utilized tool, datasets, and ML models. Section 4 provides an analysis and discussion of the obtained results. Finally, Section 5 presents the main conclusions and future work.

\section{Related Work}

In recent years, there has been a growing focus on the reliability and usability of NTA datasets for the training and validation of ML models. As the cyberattacks targeting modern IoT networks started becoming more complex, researchers started identifying limitations in how existing datasets represented that complex behaviour in a tabular data format~\cite{Thakkar2021}. Therefore, it is important to understand the conclusions and recommendations of previous work.

Some well-established studies have carefully analyzed the NTA datasets that were publicly available at the time. Kenyon et al.~\cite{kenyon_are_2020} and Ring et al.~\cite{RING2019} considered several factors, such as the quality and relevance of the captured cyberattacks, the methodology used to create and capture the traffic patterns, and the maintenance to keep it up-to-date. These studies revealed that most datasets suffer from over-summarization and have an unclear representation of benign and malicious traffic characteristics. These issues, combined with a lack of standardized methodologies for network traffic simulation, often cause the generated network flows to be incomplete and have inconsistent labels~\cite{Adeleke2022}.

Furthermore, most public NTA datasets provide a preprocessed version that loses relevant information. Even though the raw traffic is captured in PCAP files with the same format, the features made available by the original authors of each dataset are often specific to that work, and it is not specified how they were computed~\cite{rodriguez_evaluation_2022}. This results in CSV files with seemingly similar features, but that actually represent different characteristics and have missing or incompatible data, which limits the ability to combine data from different origins~\cite{GUERRA2022}.

Despite ongoing efforts to systematise the feature sets used for NTA, it is still difficult to identify which features best represent complex attack scenarios. In more recent studies, Sarhan et al.~\cite{sarhan_towards_2022} and Silva et al.~\cite{Silva2024}, evaluated different NTA datasets, including Bot-IoT and IoT-23 focused on IoT networks. Several experiments were performed to select the most relevant features for ML models to keep a good generalization while providing a more computationally efficient detection. The resulting feature sets contained mostly time-related features of active transmission rates and interpacket arrival times, but some complex cyberattack classes require the use of quantity-related features, such as sent and received bytes and packets~\cite{Vitorino2024}.

Overall, the recent literature highlights the need to perform feature engineering to better represent the characteristics and traffic patterns of more complex cyberattacks. To the best of our knowledge, no previous work has analyzed the effect that extracting new network traffic flow features from the raw network packets of the Bot-IoT, IoT-23, and CICIoT23 datasets has on the performance of ML models for multiclass cyberattack classification in IoT networks.

\section{Methodology}

This section details the traffic analysis tool, IoT datasets, and ML models used in the study which was carried out on a relatively lightweight machine equipped with a 6-core central processing unit and 16 gigabytes of random access memory. The main programming language was Python, using the following libraries: \textit{numpy} and \textit{pandas} for general data processing, and \textit{scikit-learn}, \textit{xgboost}, and \textit{lightgbm} for the implementation of the ML models.

\subsection{Traffic Analysis Tool}

HERA\footnote{\url{https://github.com/danielaapp/HERA}} was developed to simplify the creation of flow-based datasets, with or without labels, using pre-captured PCAP files, having the capabilities of a Flow Exporter and Feature Extractor~\cite{Pinto2025}. HERA processes these files through the Argus\footnote{\url{https://openargus.org/}} flow exporter and generates flows with user-defined features. The tool allowed the specification of parameters such as packet size, jitter, and the interval in seconds for flow generation, which allows flexibility in the creation of flows. To note, this flow interval definition is especially useful to have control over how much granularity the dataset presents and consequently how many flows are documented. In scenarios where a low-packet capture is generated, it might be useful to use smaller intervals, such as 1 second, so that more flows are generated to be analyzed. On the other hand, in scenarios such as Denial-of-Service (DoS) attacks, where there is a high production of network packets, using a higher interval can be beneficial to summarize flows~\cite{PintoReview2025}. Furthermore, the tool extracts features and finally allows for the labelling of data using a CSV file containing the ground truth information of the processed data.

\subsection{Datasets}

The publicly available datasets used in this study were Bot-IoT,
IoT-23,
and CICIoT23,
which are relatively recent and have been adopted by the research community for ML model benchmarks \cite{s23167191}. The following sections describe the selected datasets.

\subsubsection{Bot-IoT.}

This dataset was developed at the University of New South Wales (UNSW) 
\cite{koroniotis_towards_2019} in 2019. It was created to study IoT network activity, featuring devices such as thermostats, a smart fridge, motion-activated lights, and a weather station. The dataset included both normal IoT traffic and botnet traffic, achieved through various malware attack simulations. These attacks included probing, DoS, Distributed Denial-of-Service (DDoS), and information theft. A notable characteristic of the dataset is its class imbalance, with some attack types significantly more represented than others. To generate normal traffic, the authors used the Ostinato
tool, while IoT traffic simulation was performed using Node-RED.
Finally, flow generation was performed using the Argus tool, resulting in a dataset with a total of 47 features.

\subsubsection{IoT-23.}
 
This dataset was developed as part of the Avast Artificial Intelligence and Cybersecurity (AIC) Laboratory, funded by Avast Software in collaboration with the Czech Technical University (CTU). 
The dataset was created by Garcia et al. \cite{sebastian_garcia_2021_4743746} in 2020, to provide labelled IoT malware and benign network traffic for ML models. The latest version of IoT-23 included 8 benign captures and 20 malware captures, including a variety of malware such as Hide and Seek, Muhstik, Mirai, Hakai, Hajime, Kenjiro, Torri, Okiru, IRCBot, Trojan, and Gafgyt. The malicious traffic was created in a Raspberry Pi, while the benign traffic was captured from devices such as a Philips Hue smart LED lamp, an Amazon Echo Dot, and a Somfy Smart Door Lock. Network traffic was captured and converted into tabular flows using BRO-IDS/Zeek, and the labelling process was performed using NetflowLabeler,
a Python tool developed by CTU, resulting in a final dataset with 23 features, two of which served as labels.

\subsubsection{CICIoT23.}

This dataset was developed in 2023 by the Canadian Institute for Cybersecurity (CIC) 
\cite{neto_euclides_2023}. Similarly to the previously mentioned datasets, it was also designed to support the development of security solutions for IoT operations. The dataset includes 33 distinct attack types executed within an IoT topology comprised of 105 devices, with attack categories such as DDoS, DoS, Reconnaissance, Web-based, Brute Force, Spoofing, and Mirai attacks. These were carried out by malicious IoT devices targeting other IoT devices. The network traffic was captured using Wireshark\footnote{\url{https://www.wireshark.org/}} and stored in the PCAP format, with features extracted using the DPKT\footnote{\url{https://github.com/kbandla/dpkt}} library, 
and in CSV format with 39 computed features. Unlike previous datasets, CICIoT23 organized its ``flows'' based on fixed-size windows with a certain amount of packets, contrary to the other datasets, which follow the definition of NetFlow. Most attack and benign traffic are grouped in windows of 10 packets, while high-volume attacks, such as DDoS and Command Injection, use 100 to compensate for their larger traffic generation.

\subsection{Data Preprocessing}

\begin{figure}[ht]
    \centering
    \includegraphics[width=0.99\linewidth]{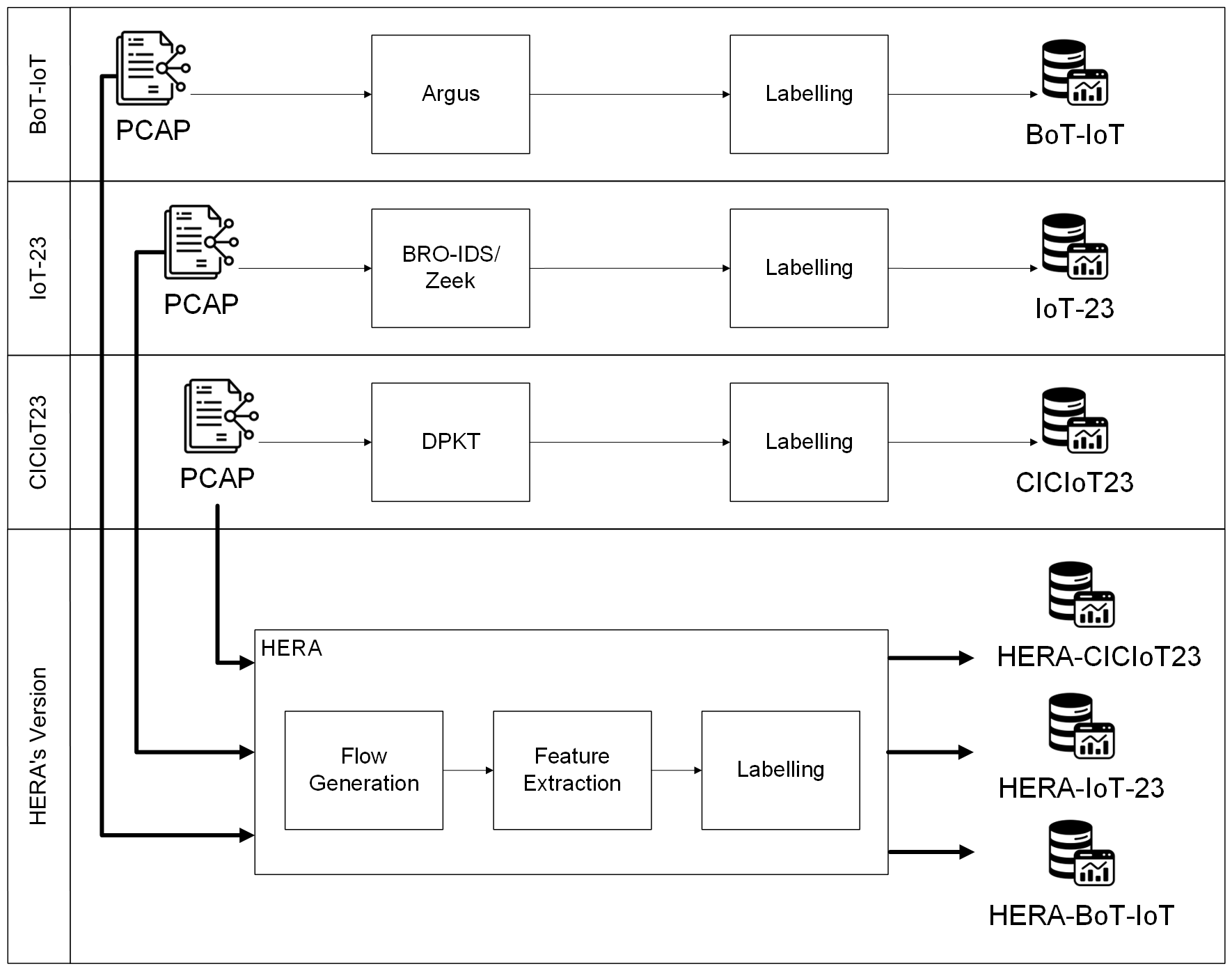}
    \caption{Workflow of dataset creation and data preprocessing.}
    \label{fig:workflow}
\end{figure}

In this study, the HERA tool was applied to the raw PCAP files from multiple NTA datasets to produce new CSV versions with network flows that are compatible with each other, following a 60-second time interval. The generated flows were labelled according to the documentation and ground-truth labels made available by the original authors of each dataset: Bot-IoT included ground-truth files, IoT-23 provided documentation and partial labelling files, and CICIoT23 segregated the PCAP files directly into benign and malicious traffic. In addition to the computed features for ML models, the new CSV versions also included the attack start and end times and source and destination Internet Protocol (IP) addresses. Fig. \ref{fig:workflow} provides an overview of the data preprocessing.

The following subsections describe the features that were present in the original dataset and their HERA versions, as well as the methodology used to label the data. It is important to note that the three HERA versions emulate as closely as possible the original features, to enable a more trustworthy comparison of how using a different tool impacts the generated flows.

\subsubsection{Bot-IoT.}

Since this dataset uses the same flow exporter as HERA, Argus, the extracted features matched those originally used. A single label representing the attack type was used for multiclass classification, and HERA introduced a ``FlowID'', which was composed of the source and destination IP addresses, ports, and the protocol of the flow, and a ``Rank'', which indicated the order of flows within each file. In contrast, Bot-IoT used only the ``pkSeqID'' feature, which served a similar purpose but operated indistinctively across all dataset files rather than being exclusive to individual files as in HERA.

The tool was configured to compute multiple useful characteristics of each network flow: start and last time of the flow (``StartTime'' and ``LastTime''), the transaction protocol (``Proto''), source and destination IP address and port number (``SrcAddr'', ``Sport'', ``DstAddr'' and ``Dport''), the flow state flags (``Flgs), total, source and destination transaction packet and byte counts.  (``TotPkts'', ``SrcPkts'', ``DstPkts'', ``TotBytes'', ``SrcBytes'' and ``DstBytes''), the transaction state (``State''), the sequence number (``Seq''), the record total duration (``Dur''), the average and standard deviation of the duration of the records in the flows (``Mean'' and ``StdDev''), the source and destination medium access control (MAC) addresses (``SrcMac'' and ``DstMac''), the total, the minimum and maximum duration of the records in the flows (``Sum'', ``Min'' and ``Max''), the source and destination organizationally unique identifier (OUI) portion of the MAC address (``SrcOui'' and ``DstOui''), the source and destination IP address country code (``sCo'' and ``dCo''), and the total, source and destination packets per second rates (``Rate'', ``SrcRate'' and ``DstRate'').

Nonetheless, not all of these characteristics were useful features for the training of ML models. Table \ref{tab_features_botiot} presents the feature sets that were used in the final evaluation. It is important to note that even though these features are equivalent to the original Bot-IoT version, they were computed for packet sequences with different flow intervals, so they ultimately represent different network flows. Additionally, the ``Rate'' feature was used to also include information about the total packets transmitted per second.


\begin{table}[ht]
    \centering
    \caption{Utilised features for Bot-IoT dataset.}
    \label{tab_features_botiot}
    \begin{tabular}{ >{\centering\arraybackslash}p{3.5cm} >{\centering\arraybackslash}p{3.5cm} }
        \hline
        \centering\textbf{Original Features} & \textbf{HERA Features} \\ \hline
        Sport & Sport \\
        Dport & Dport \\
        Dur & Dur \\
        SrcBytes & SrcBytes \\
        DstBytes & DstBytes \\
        SrcPkts & SrcPkts \\
        DstPkts & DstPkts \\
        Mean & Mean \\
        StdDev & StdDev \\
        Max & Max \\
        Min & Min \\
        - & Rate \\ \hline
    \end{tabular}
\end{table}

For the final evaluation, all attack types were fully represented, except for UDP and Transmission Control Protocol (TCP) DoS and DDoS, in which only a small portion of around 1/10th was used due to the large quantity of repeated network flows generated in these attacks. Since the original and HERA versions contained different network flows, their total quantities and flow distribution among the CSV files also varies. Since the HERA version used a 60-second flow interval, which is higher than the 5-second flow interval of the original version, there could be a reduction in the number of generated flows. Nonetheless, this was counteracted by the more in-depth analysis of the network connections performed by HERA, which produced approximately two million flows more than the original, as presented in Table \ref{tab_flows_botiot}.

An additional aspect is that, for the original dataset, the authors performed an arbitrary division of the entire dataset to contain around 1 million flows in each CSV file. Since the HERA version was created based on the division of the PCAP files, instead of the arbitrary division of the CSV files, it contains a very slightly bigger quantity of packets for some attacks. Even though it is not a big discrepancy time-wise, the attacks that generated millions of packets seem to contribute heavily to the difference in the number of generated flows.

\begin{table}[ht]
    \centering
    \caption{Obtained flows for Bot-IoT dataset.}
    \label{tab_flows_botiot}
    \begin{tabular}{>{\centering\arraybackslash}p{3.5cm} >{\centering\arraybackslash}p{3.5cm} >{\centering\arraybackslash}p{3.5cm}}
        \hline
        \textbf{Class Name} & \textbf{HERA Flows} & \textbf{Original Flows} \\
        \hline
        DoS TCP & 3.509.418 & 2.316.439 \\
        DoS UDP & 2.883.703 & 3.660.048 \\
        DDoS TCP & 2.243.706 & 2.548.154 \\
        DDoS UDP & 2.097.220 & 1.965.658 \\
        Benign & 1.601.930 & 7.441 \\
        Service Scan & 1.472.789 & 1.463.364 \\
        OS Scan & 360.511 & 358.275 \\
        DoS HTTP & 35.505 & 29.706 \\
        DDoS HTTP & 25.708 & 19.771 \\
        Keylogging & 1.483 & 1.469 \\
        Data Exfiltration & 139 & 118 \\
        \hline
        \textbf{Total} & 14.232.112 & 12.370.443 \\
        \hline
    \end{tabular}
\end{table}

\subsubsection{IoT-23.}

The original dataset provides a binary label, for benign or malicious traffic, as well as a more detailed label specifying the utilized malware. To label the HERA’s version, the adopted approach consisted of an analysis of the combinations of IPs, ports, and protocols identified as malicious.

Regarding the features present in IoT-23, as they were originally computed using the BRO-IDS/Zeek tool, an exact match could not be extracted. Nonetheless, since HERA relies on similar computations, the selected feature set resembled as closely as possible the one CTU originally used. This resulted in some features that had also been used for Bot-IoT: ``FlowID'', ``Rank'', ``StartTime'', ``LastTime'', ``Proto'', ``SrcAddr'', ``Sport'', ``DstAddr'', ``Dport'', ``Dur'', ``TotPkts'', ``SrcPkts'', ``DstPkts'', ``TotBytes'', ``SrcBytes'', and ``DstBytes''. Four additional features were also extracted: the source and destination interpacket arrival times (``SIntPkt'' and ``DIntPkt''), and the source and destination bytes missing in the data stream (``SrcGap'' and ``DstGap''). Table \ref{tab_features_iot23} presents the final feature sets. 


\begin{table}[ht]
    \centering
    \caption{Utilised features for IoT-23 dataset.}
    \label{tab_features_iot23}
    \begin{tabular}{ >{\centering\arraybackslash}p{3.5cm} >{\centering\arraybackslash}p{3.5cm} }
        \hline
        \textbf{Original Features} & \textbf{HERA Features} \\ \hline
        id.orig-p & Sport \\
        id.resp-p & Dport \\
        duration & Dur \\
        orig-bytes & SrcBytes \\
        resp-bytes & DstBytes \\
        orig-pkts & SrcPkts \\
        resp-pkts & DstPkts \\
        - & SIntPkt \\
        - & DIntPkt \\
        - & SrcGap \\
        - & DstGap \\ \hline
    \end{tabular}
\end{table}

Furthermore, due to the large quantity of repeated traffic in IoT-23, only a subset with the attacks Mirai, Gafgyt, Hajime, Hide\&Seek and Muhstik was used in both original and HERA versions. Table \ref{tab_flows_iot23} presents the discrepancy in the number of generated flows in the two dataset versions. These quantities were obtained from the exact same PCAP files, with the only difference being the tool used for traffic aggregation and flow generation. The reduced quantity of the HERA version compared to the original version evidences that there are significant differences in the way that the utilized tools aggregate packet sequences into flows.

\begin{table}[ht]
    \centering
    \caption{Obtained flows for IoT-23 dataset.}
    \label{tab_flows_iot23}
    \begin{tabular}{>{\centering\arraybackslash}p{3.5cm} >{\centering\arraybackslash}p{3.5cm} >{\centering\arraybackslash}p{3.5cm}}
        \hline
        \textbf{Class Name} & \textbf{HERA Flows} & \textbf{Original Flows} \\
        \hline
        Mirai & 3.434.426 & 20.263.372 \\
        Gafgyt & 1.059.236 & 3.578.552 \\
        Hajime & 1.045.987 & 6.355.745 \\
        Hide\&Seek & 610.240 & 539.473 \\
        Muhstik & 85.213 & 151.567 \\
        Benign & 68.365 & 518.458    \\
        \hline
        \textbf{Total} & 6.303.467 & 31.407.167 \\
        \hline
    \end{tabular}
\end{table}

\subsubsection{CICIoT23.}

Regarding the HERA’s version of this dataset, it was considered that the only traffic present in each capture is exclusively the one identified in its name. The features were again extracted to represent as closely as possible the same characteristic as the original ones: ``FlowID'', ``Rank'', ``StartTime'', ``LastTime'', ``Proto'', ``SrcAddr'', ``Sport'', ``DstAddr'', ``Dport'', ``Dur'', ``Mean'', ``StdDev'', ``Min'', ``Max'', ``Rate'', ``SrcRate'', ``DstRate'', ``SIntPkt'' and ``DIntPkt'', described in the previous subsections. Table \ref{tab_features_ciciot23} presents the final feature sets.


\begin{table}[ht]
    \centering
    \caption{Utilised features for CICIoT23 dataset.}
    \label{tab_features_ciciot23}
    \begin{tabular}{ >{\centering\arraybackslash}p{3.5cm} >{\centering\arraybackslash}p{3.5cm} }
        \hline
        \textbf{Original Features} & \textbf{HERA Features} \\ \hline
        - & Sport \\
        - & Dport \\
        dur & Dur \\
        max & Max \\
        min & Min \\
        rate & Rate \\
        srate & - \\
        drate & - \\
        iat & - \\
        avg & Mean \\
        std & StdDev \\ \hline
    \end{tabular}
\end{table}

The discrepancy between the different methodologies used to preprocess the datasets, generating different amounts of network flows for the same network packets of the original PCAP files, is presented once again in Table \ref{tab_flows_ciciot23}, for the CICIoT23 dataset. In contrast with IoT-23, the HERA version of CICIoT23 was the one to contain an increased amount of flows, dividing both benign and malicious traffic into a greater quantity of packet sequences with different flow intervals, which can impact the training process of ML classification models.

\begin{table}[ht]
    \centering
    \caption{Obtained flows for CICIoT23 dataset.}
    \label{tab_flows_ciciot23}
    \begin{tabular}{>{\centering\arraybackslash}p{4.5cm} >{\centering\arraybackslash}p{3cm} >{\centering\arraybackslash}p{3cm}}
        \hline
        \textbf{Class Name} & \textbf{HERA Flows} & \textbf{Original Flows} \\
        \hline
        DDoS SYN Flood & 7.292.680 & 934.867 \\
        DDoS ICMP Flood & 6.754.146 & 1.071.969 \\
        DDoS RSTFIN Flood & 6.191.448 & 863.517 \\
        DDoS TCP Flood & 6.109.125 & 893.650 \\
        DDoS PSHACK Flood & 6.084.437 & 883.330 \\
        DoS SYN Flood & 5.602.405 & 995.726 \\
        DoS TCP Flood & 5.223.704 & 822.825 \\
        Mirai Greeth Flood & 5.186.402 & 135.890 \\
        Mirai Greip Flood & 5.141.002 & 117.281 \\
        DDoS Synonymous IP Flood & 4.702.200 & 931.987 \\
        DDoS ICMP Fragmentation & 3.596.210 & 85.610 \\
        DDoS ACK Fragmentation & 3.479.543 & 79.749 \\
        DoS UDP Flood & 3.000.282 & 754.763 \\
        DoS HTTP Flood & 892.500 & 71.861 \\
        Recon Host Discovery & 722.968 & 134.378 \\
        Benign & 595.785 & 1.098.191 \\
        Vulnerability Scan & 547.506 & 373.351 \\
        DDoS HTTP Flood & 473.119 & 28.790 \\
        Recon Port Scan & 233.295 & 82.284 \\
        Recon OS Scan & 208.315 & 98.259 \\
        DDoS Slowloris & 175.604 & 23.426 \\
        DDoS UDP Flood & 158.087 & 880.769 \\
        DNS Spoofing & 130.163 & 178.898 \\
        MITM ARP Spoofing & 128.671 & 307.560 \\
        DDoS UDP Fragmentation & 107.501 & 83.523 \\
        Mirai UDP Plain & 37.682 & 124.787 \\
        Dictionary Brute Force & 12.888 & 13.064 \\
        SQL Injection & 7.612 & 5.245 \\
        Command Injection & 6.576 & 5.409 \\
        Backdoor Malware & 4.452 & 3.218 \\
        Recon Ping Sweep & 2.808 & 2.262 \\       
        \hline
        \textbf{Total} & 72.809.116 & 12.086.439 \\
        \hline
    \end{tabular}
\end{table}

\subsection{Model Fine-tuning}

\setlength{\tabcolsep}{10pt}

Since decision tree ensembles obtained promising results for NTA and cyberattack classification in the surveyed previous work, the selected ensembles for this study were RF, XGB, and LGBM. Independent models were trained and fine-tuned to each dataset version, employing 5-fold cross-validation combined with a grid search of well-performing hyperparameters for each model. After the fine-tuning, each model was retrained with a complete training set and the final evaluation was performed with the holdout set of each dataset version.

Since the network traffic of real-world IoT networks is not always sent in the same quantities and in same time intervals, the considered NTA datasets contain imbalanced data with different class proportions. Therefore, the macro-averaged F1-Score was selected as the validation metric. In each iteration of the cross-validation, a model was trained with 4/5 of a training set and validated with the remaining 1/5, using well-established hyperparameters for multiclass cyberattack classification to attempt to maximize this metric.

\subsubsection{Random Forest.}

RF is a supervised ensemble that is considered a bagging algorithm. It performs bootstrap aggregation using random subsets of features of a dataset, creating many decision trees based on the concept of the wisdom of the crowd: the collective decisions of multiple classifiers will be better than the decisions of just one.

It was configured to use the Gini Impurity criterion to measure the quality of the possible node splits, and the maximum depth of a tree and the minimum number of samples in a leaf node were optimized for each training set, resulting in different values for each dataset.
Table \ref{tab_model_rf} summarizes the configuration.

\begin{table}[ht]
\centering
\caption{Summary of RF configuration.}
\label{tab_model_rf}
\begin{tabular}{cc}
\hline
\textbf{Hyperparameter} & \textbf{Value} \\
\hline
Criterion & Gini impurity \\
No. of estimators & 100 \\
Max. depth of a tree & 4 to 16 \\
Min. samples in a leaf & 2 to 4 \\
\hline
\end{tabular}
\end{table}

\vspace{-5pt}

\subsubsection{Extreme Gradient Boosting.}

XGB is a gradient boosting algorithm that also uses a supervised ensemble of decision trees. It relies on a level-wise growth strategy to split tree nodes level by level, seeking to minimize a loss function during the boosting process.

The Cross-Entropy loss was used for multiclass classification, employing the Histogram method to choose the best node splits by computing fast histogram-based approximations. The main hyperparameter of this ensemble is the learning rate of the boosting process, which controls how quickly it adapts its trees to the training data but is also affected by the values of several other hyperparameters.
Table \ref{tab_model_xgb} summarizes the configuration.

\begin{table}[ht]
\centering
\caption{Summary of XGB configuration.}
\label{tab_model_xgb}
\begin{tabular}{cc}
\hline
\textbf{Hyperparameter} & \textbf{Value} \\
\hline
Method & Histogram \\
Loss function & Cross-entropy \\
No. of estimators & 100 \\
Learning rate & 0.05 to 0.25 \\
Max. depth of a tree & 4 to 16 \\
Min. loss reduction & 0.01 \\
Feature subsample & 0.7 to 0.9 \\
\hline
\end{tabular}
\end{table}

\subsubsection{Light Gradient Boosting Machine.}

LGBM also uses a supervised tree ensemble to perform gradient boosting, but it employs a different approach than XGB. It relies on a leaf-wise strategy, following a best-first approach to directly split the leaf node with the maximum loss reduction at any level of a tree.

The main advantage of this ensemble is the use of Gradient-based One-Side Sampling (GOSS) to build the decision trees, which is computationally lighter than previous methods and therefore provides a faster-boosting process. The Cross-Entropy loss was also used, and the learning rate was kept at small values to avoid fast convergences to suboptimal solutions in the imbalanced datasets.
Table \ref{tab_model_lgbm} summarizes the configuration.

\begin{table}[ht]
\centering
\caption{Summary of LGBM configuration.}
\label{tab_model_lgbm}
\begin{tabular}{cc}
\hline
\textbf{Hyperparameter} & \textbf{Value} \\
\hline
Method & GOSS \\
Loss function & Cross-entropy \\
No. of estimators & 100 \\
Learning rate & 0.05 to 0.15 \\
Max. leaves in a tree & 5 to 10 \\
Min. loss reduction & 0.01 \\
Min. samples in a leaf & 2 to 4 \\
Feature subsample & 0.8 to 0.9 \\
\hline
\end{tabular}
\end{table}

\section{Results and Discussion}

\setlength{\tabcolsep}{5pt}

This section presents a comparative analysis of the evaluation results of the RF, XGB, and LGBM models for the Bot-IoT, IoT-23, and CICIoT23 datasets, considering the standard evaluation metrics for imbalanced multiclass classification tasks: accuracy, macro-averaged precision, recall, and F1-Score.

\subsection{Bot-IoT}

The models trained with the original version of the Bot-IoT dataset obtained very high accuracy scores, from 97.88\% to 99.32\%. Both RF and XGB were able to reach macro-averaged F1-Scores of approximately 99\%, which indicates that each cyberattack class of this dataset can be easily detected by ML models. Nonetheless, despite the fine-tuned LGBM obtaining 97.88\% accuracy, it did not detect some minority cyberattack classes and only reached an F1-Score of 62.90\%. This suggests that the original network flows of some of the cyberattack classes of this dataset may have some discrepancies or biases that affect the learning process of certain decision tree ensembles.

By using the HERA tool to extract new network flows from the original PCAP files, a novel version of the Bot-IoT dataset was created and new models were trained. Even though the F1-Scores of RF and XGB had a small decrease of approximately 2.7\% and 1.0\%, the precision and recall of LGBM were greatly increased, leading to an F1-Score 33.9\% higher. Since the features of both versions were equivalent, these results show that changing the tool used to aggregate raw packets and extract network traffic flow features impacted the generalization of the models. Table \ref{tab_results_botiot} summarizes the obtained results for each version, with "M." representing "Macro-averaged".

\begin{table*}[ht]
\centering
\caption{Obtained results for Bot-IoT dataset.}
\label{tab_results_botiot}
\begin{tabular}{cccccc}
\hline
\textbf{Dataset} & \textbf{Model} & \textbf{Accuracy} & \textbf{M. Precision} & \textbf{M. Recall} & \textbf{M. F1-Score} \\
\hline
\multirow{3}{4em}{\hfil Original \\ \hfil Flows}
& RF & 99.22 & 99.29 & 98.99 & 99.14 \\
& XGB & 99.32 & 98.88 & 98.72 & 98.80 \\
& LGBM & 97.88 & 62.55 & 63.36 & 62.90 \\
\hline
\multirow{3}{4em}{\hfil HERA \\ \hfil Flows}
& RF & 97.29 & 98.08 & 94.96 & 96.42 \\
& XGB & 97.77 & 98.33 & 97.34 & 97.82 \\
& LGBM & 96.84 & 97.42 & 96.13 & 96.75 \\
\hline
\end{tabular}
\end{table*}

\subsection{IoT-23}

In contrast with Bot-IoT, the original version of the IoT-23 dataset did not provide good results for multiclass classification. The three models only had an accuracy near 64\%, and despite reaching a relatively high precision, their low recall led to F1-Scores of approximately 42\% for the fine-tuned RF, 40\% for XGB, and 31\% for LGBM. This indicates that there was a relatively low number of false positives across cyberattack classes, as the majority of network flows predicted to belong to a certain cyberattack were actually of that class, but there was a relatively high number of false negatives, with many malicious flows not being detected by the models.

Since the original flows had many missing values in categorical features and sparse data in numerical features, recreating these features and generating more complete flows with the HERA tool significantly improved the generalization of the three ML models. The models trained and fine-tuned to the new version of the IoT-23 dataset achieved accuracy scores of over 99.8\% and F1-Scores of over 99.2\%. This highlights the benefits of directly preprocessing the raw network traffic of PCAP files to compute more relevant features, instead of just using the incomplete flows of the commonly available CSV files. Table \ref{tab_results_iot23} summarizes the obtained results for each version.

\begin{table*}[ht]
\centering
\caption{Obtained results for IoT-23 dataset.}
\label{tab_results_iot23}
\begin{tabular}{cccccc}
\hline
\textbf{Dataset} & \textbf{Model} & \textbf{Accuracy} & \textbf{M. Precision} & \textbf{M. Recall} & \textbf{M. F1-Score} \\
\hline
\multirow{3}{4em}{\hfil Original \\ \hfil Flows}
& RF & 64.17 & 77.45 & 38.08 & 42.21 \\
& XGB & 63.96 & 61.72 & 36.32 & 39.75 \\
& LGBM & 62.75 & 42.33 & 31.57 & 31.06 \\
\hline
\multirow{3}{4em}{\hfil HERA \\ \hfil Flows}
& RF & 99.92 & 99.22 & 99.59 & 99.40 \\
& XGB & 99.88 & 99.21 & 99.26 & 99.23 \\
& LGBM & 99.89 & 98.96 & 99.48 & 99.21 \\
\hline
\end{tabular}
\end{table*}

\subsection{CICIoT23}

The models trained with the most recent dataset, CICIoT23, reached reasonably good results, with accuracy scores from 73\% to 76\% and macro-averaged F1-Scores from 54\% to 59\%, with higher precision than recall. These results, lower than the older Bot-IoT, indicate that decision tree ensembles do not perform as well in multiclass classification of more recent cyberattacks, possibly due to the difficulty in representing more complex network traffic patterns in a simple tabular data format suitable for ML models.

The HERA version of the CICIoT23 dataset enabled an increase of over 20\% in accuracy in all models fine-tuned to the new network flows. Even though the F1-Scores of RF, XGB, and LGBM were only increased to 64.61\%, 60.13\%, and 57.43\%, respectively. These are not optimal results, but all bagging and boosting tree ensembles still achieved a good generalization, considering that this multiclass classification task includes minority cyberattack classes that are harder to detect. Table \ref{tab_results_ciciot23} summarizes the obtained results.

\begin{table*}[ht]
\centering
\caption{Obtained results for CICIoT23 dataset.}
\label{tab_results_ciciot23}
\begin{tabular}{cccccc}
\hline
\textbf{Dataset} & \textbf{Model} & \textbf{Accuracy} & \textbf{M. Precision} & \textbf{M. Recall} & \textbf{M. F1-Score} \\
\hline
\multirow{3}{4em}{\hfil Original \\ \hfil Flows}
& RF & 76.41 & 74.17 & 57.82 & 59.49 \\
& XGB & 76.46 & 79.21 & 56.96 & 58.44 \\
& LGBM & 73.61 & 57.59 & 54.83 & 54.40 \\
\hline
\multirow{3}{4em}{\hfil HERA \\ \hfil Flows}
& RF & 96.46 & 67.47 & 62.83 & 64.61 \\
& XGB & 94.95 & 80.04 & 57.21 & 60.13 \\
& LGBM & 94.38 & 65.50 & 55.23 & 57.43 \\
\hline
\end{tabular}
\end{table*}

\section{Conclusions}

This work addressed the impact that network traffic flow exporters can have on the robustness of ML models for the detection and classification of cyberattacks in IoT networks. The raw PCAP files of three NTA datasets were analyzed with the HERA tool, generating new labelled flows and extracting features that were compared with the original versions. To assess the usefulness of these new flows for the training of ML-based intelligent detection systems, multiple ML models were fine-tuned and evaluated with these datasets.

The considered decision tree ensembles, RF, XGB, and LGBM, obtained good results in an imbalanced multiclass classification task with the Bot-IoT, IoT-23, and CICIoT23 datasets. Even though no improvement was noticed on the older dataset, a better generalization was achieved in the HERA versions of the two most recent datasets, just by changing the way that packet sequences are aggregated into flows. The models had improvements of up to 37\% in accuracy and, more notably, up to 68\% in macro-averaged F1-Score, reaching scores above 99\% in one of the recent NTA datasets.

Overall, the obtained results indicate that directly analysing and preprocessing the raw traffic of PCAP files, instead of just using the network flows of existing CSV files, enables the computation of more relevant features to train more robust ML models. It is important to continue improving feature extraction and feature selection processes from raw traffic, as it contributes to the representation of the increasingly complex network traffic patterns of cyberattacks in a simple tabular data format suitable for ML models.

To continue this work, more flow exporters and feature extraction tools should be explored, and a more in-depth analysis should be performed to inspect how these tools compute each feature, as it can significantly impact the generalization and robustness ML models. As more NTA datasets start becoming publicly available, it will become essential to define standard feature sets for different types of network topologies, to make datasets from different origins more compatible with each other and enable a trustworthy evaluation and comparison of the ML models used in cybersecurity solutions.

\section*{\uppercase{Acknowledgements}}
This work was supported by the SAFE project, which has received funding from the European Cybersecurity Competence Centre under grant agreement 101190370. This work also received funding from UIDB/00760/2020.
\bibliographystyle{splncs04}
\bibliography{bibliography.bib}
\end{document}